\documentclass[preprint,12pt,3p]{elsarticle}

\usepackage{matlab-prettifier}
\usepackage{graphicx}
\usepackage{amsmath}
\usepackage{parskip}
\usepackage{hyperref}
\usepackage{adjustbox}
\usepackage{setspace}
\hypersetup{ 
backref=true, 
bookmarks=true, 
unicode=false, 
pdftoolbar=true, 
pdfmenubar=true, 
pdffitwindow=false, 
pdfstartview={FitH}, 
pdftitle={}, 
pdfauthor={}, 
pdfsubject={}, 
pdfkeywords={}{}, 
pdfnewwindow=true, 
colorlinks=true, 
linkcolor=BlueViolet, 
citecolor=maroon, 
urlcolor=blue,
}
\usepackage{comment}
\usepackage{subfig}
\usepackage{caption}
\usepackage{amssymb}
\usepackage{gensymb}

\usepackage{natbib}
\biboptions{comma,round}
\setcitestyle{authoryear}

\begin{document}
\begin{frontmatter}

\title{Climate Determinism Reborn}

\author[label2,label3,label4,label5,label6,label7]{Richard S.J. Tol\corref{cor1}\fnref{label9}}
\address[label2]{Department of Economics, University of Sussex, Falmer, United Kingdom}
\address[label3]{Institute for Environmental Studies, Vrije Universiteit, Amsterdam, The Netherlands}
\address[label4]{Department of Spatial Economics, Vrije Universiteit, Amsterdam, The Netherlands}
\address[label5]{Tinbergen Institute, Amsterdam, The Netherlands}
\address[label6]{CESifo, Munich, Germany}
\address[label7]{Payne Institute for Public Policy, Colorado School of Mines, Golden, CO, United States of America}

\cortext[cor1]{Jubilee Building, BN1 9SL, UK}
\fntext[label9]{Thank you.}

\ead{r.tol@sussex.ac.uk}
\ead[url]{http://www.ae-info.org/ae/Member/Tol\_Richard}

\begin{abstract}
Environmental determinism in the past followed from the belief that the gods bestowed political power and the best possible weather on the sponsors of early scholars. Although later discredited in academia because of the associations with racism and the lack of support for any monocausal explanation of history, environmental determinism in popular culture has morphed into the unfounded idea of catastrophic climate change. Although a handful of papers in the literature on the economic impact of climate change appear to support this concern, closer inspection reveals severe methodological and conceptual issues with the analyses. Climate policy will not dominate major economies, because politicians will pull back before it does, but small economies may be overwhelmed by the export of carbon credits.
\end{abstract}

\end{frontmatter}

\newpage
\section{Introduction}
\label{sc:intro}
People used to believe silly things. The Sun revolves around the Earth. Bad air causes malaria. The free market is best. Climate determinism is one of those things. Or is it? The idea that the natural environment determines the human condition was once widespread. After a century-long lapse, it has returned.

Attempts to explain or justify the political and military power of the benefactors of early scholars coloured their views on the role of climate and geography in economic development. A perfect climate is one of the divine blessings on the hegemon and a reason for foreigners' lack of power and civility. Such theories persisted well into the 19th century.

The neo-Malthusian views of the Club of Rome revived environmental determinism\textemdash not as an explanation for the success of the powers that be, but as a prediction for their demise. Over time, the environmental movement shifted its attention away from resource scarcity and local pollution to global climate change, which has duly become the destroyer of worlds.

I here push back against these arguments. Climate and climate change were never the single cause for historical developments and their influence waned with the importance of agriculture and the increasing ability to shield ourselves from the environment and shape it to suit our needs. I argue below that recent claims that climate change would devastate economies are based on subpar applications of econometric methods.

The paper proceeds as follows. Section \ref{sc:past} reviews the history of environmental determinism and the counterarguments. Section \ref{sc:present} surveys the climate determinism of today, touching on popular authors, but focusing on a relatively small group of academic economists. Section \ref{sc:future} changes gear, discussing how, if the current political rhetoric will come to pass, climate \emph{policy} would dominate future economies and societies for the rest of the century. Section \ref{sc:conclude} concludes.

\section{Climate determinism past}
\label{sc:past}
Research requires a surplus and is therefore concentrated near power. Before the Enlightenment, learned people argued that divine beings shaped the environment and smiled on the powerful. It follows that, through the ages, scholars maintained that the climate of the hegemon was the best the gods could do.

For instance, Hippocrates \href{http://classics.mit.edu/Hippocrates/airwatpl.23.23.html}{(400 BCE, On Airs, Waters, Places, 23)} wrote
\begin{quote}
The other races in Europe differ from one another, both as to stature and shape, owing to the changes of the seasons, which are very great and frequent, and because the heat is strong, the winters severe, and there are frequent rains, and again protracted droughts, and winds, from which many and diversified changes are induced. These changes are likely to have an effect upon generation in the coagulation of the semen, as this process cannot be the same in summer as in winter, nor in rainy as in dry weather; wherefore, I think, that the figures of Europeans differ more than those of Asiatics; and they differ very much from one another as to stature in the same city; for vitiations of the semen occur in its coagulation more frequently during frequent changes of the seasons, than where they are alike and equable. And the same may be said of their dispositions, for the wild, and unsociable, and the passionate occur in such a constitution; for frequent excitement of the mind induces wildness, and extinguishes sociableness and mildness of disposition, and therefore I think the inhabitants of Europe more courageous than those of Asia; for a climate which is always the same induces indolence, but a changeable climate, laborious exertions both of body and mind; and from rest and indolence cowardice is engendered, and from laborious exertions and pains, courage. On this account the inhabitants of Europe are than the Asiatics, and also owing to their institutions, because they are not governed by kings like the latter, for where men are governed by kings there they must be very cowardly, as I have stated before; for their souls are enslaved, and they will not willingly, or readily undergo dangers in order to promote the power of another; but those that are free undertake dangers on their own account, and not for the sake of others; they court hazard and go out to meet it, for they themselves bear off the rewards of victory, and thus their institutions contribute not a little to their courage.
\end{quote}
Hippocrates argues that there are all sorts of things wrong with non-Greek Europeans and Asiatics\textemdash because of where they live\textemdash implying that the climate in Greece is responsible for the superiority of him and his countrymen.

Aristotle \href{http://classics.mit.edu/Aristotle/politics.7.seven.html}{(350 BCE, Politics, 7 VII)} agrees, writing
\begin{quote}
    Those who live in a cold climate and in Europe are full of spirit, but wanting in intelligence and skill; and therefore they retain comparative freedom, but have no political organization, and are incapable of ruling over others. Whereas the natives of Asia are intelligent and inventive, but they are wanting in spirit, and therefore they are always in a state of subjection and slavery. But the Hellenic race, which is situated between them, is likewise intermediate in character, being high-spirited and also intelligent. Hence it continues free, and is the best-governed of any nation, and, if it could be formed into one state, would be able to rule the world.
\end{quote}
Like Hippocrates, Aristotle argued that his own people were superior by virtue of their climate. He correctly predicted that his student, Alexander the Great, would rule the world\textemdash but he was incorrect about the yet-to-be-built empires in Europe and Asia. He somehow overlooked the Asian empires that preceded his time, as well as the contemporaneous Persian one.

The ancient Greeks were not alone in their environmental determinism. In an important, early Chinese text, Guan Zhong\href{https://books.google.co.uk/books?id=Yu7eDwAAQBAJ\&pg=PA106&lpg=PA106&dq=guanzi+greedy,+uncouth,+and+warlike\&source=bl\&ots=6eLMGGPbXp\&sig=ACfU3U1ckHgtbfjTESsSPRnWIziRpLRvIQ\&hl=en\&sa=X\&ved=2ahUKEwiDqpqljZrqAhUSZcAKHWNkCCIQ6AEwAHoECAkQAQ#v=onepage\&q=guanzi\%20greedy\%2C\%20uncouth\%2C\%20and\%20warlike\&f=false}{(780 BCE, Guanzi, XIV 39)} wrote
\begin{quote}
    What is water? It is the root of all things and the ancestral hall of all life. It is that from which beauty and ugliness, worthiness and unworthiness, stupidity and giftedness are produced.
    
    How do we know this to be so? Now the water of Qi is forceful, swift and twisting. Therefore its people are greedy, uncouth and warlike. The water of Chu is gentle, yielding, and pure. Therefore its people are lighthearted, resolute, and sure of themselves. The water of Yue is turbid, sluggish, and soaks the land. Therefore its people are stupid, disease ridden, and filthy. The water of Qin is thick like gruel and stagnant. It is obstructed, choked with silt, and wanders in confusion free of its banks. Therefore its people are greedy, violent, and deceptive, and they like to meddle in affairs. The water of Jin is bitter, harsh, and polluted. Therefore its people are flattering and deceitful, cunning and profit seeking. The water of Yan collects in low places and is weak. It sinks into the ground, is clogged, and wanders in confusion free of its banks. Therefore its people are stupid, idiotic, and given to divination. They treat disease lightly and die readily. The water of Song is light, strong, and pure. Therefore its people are simple and at ease with themselves, and they like things to be done in the correct way. For this reason, the sages' transformation in the world lay in understanding water.
    
    Now, when the water is unadulterated, people's hearts will be correct, they have no desire to be corrupt. When people's hearts are at ease, their conduct will never be depraved. For this reason, the sages' bringing good order to the world did not lie in preaching to every person or persuading every household, but in taking water as their central concern.
\end{quote}
Guan Zhong says that you are like the nearest river. He argues that a wise ruler improves water courses and thus the character of his subjects. He so justifies the elite in a hydraulic society, whose main role was to provide public goods in water management.

The idea of environmental determinism lingered. Ibn Khaldun \href{http://www.muslimphilosophy.com/ik/Muqaddimah/Chapter1/Ch_1_04.htm}{(1377, Muqaddimah 1 4)}
\begin{quote}
    We have seen that Negroes are in general characterized by levity, excitability, and great emotionalism. They are found eager to dance whenever they hear a melody. They are everywhere described as stupid. The real reason for these (opinions) is that, as has been shown by philosophers in the proper place, joy and gladness are due to expansion and diffusion of the animal spirit. Sadness is due to the opposite, namely, contraction and concentration of the animal spirit. It has been shown that heat expands and rarefies air and vapors and increases their quantity. [...]

    Now, Negroes live in the hot zone (of the earth). Heat dominates their temperament and formation. Therefore, they have in their spirits an amount of heat corresponding to that in their bodies and that of the zone in which they live. In comparison with the spirits of the inhabitants of the fourth zone,\footnote{Ibn Khaldun split the world into seven climate zones. The fourth zone, the middle one, had the optimal climate. Other zones were too hot or too cold. Ibn Khaldun lived in, where else, the optimal climate.} theirs are hotter and, consequently, more expanded. As a result, they are more quickly moved to joy and gladness, and they are merrier. Excitability is the direct consequence.

    In the same way, the inhabitants of coastal regions are somewhat similar to the inhabitants of the south. The air in which they live is very much hotter because of the reflection of the light and the rays of (the sun from) the surface of the sea. Therefore, their share in the qualities resulting from heat, that is, joy and levity, is larger than that of the (inhabitants of) cold and hilly or mountainous countries. To a degree, this may be observed in the inhabitants of the Jarid in the third zone. The heat is abundant in it and in the air there, since it lies south of the coastal plains and hills. Another example is furnished by the Egyptians. Egypt lies at about the same latitude as the Jarid. The Egyptians are dominated by joyfulness, levity, and disregard for the future. They store no provisions of food, neither for a month nor a year ahead, but purchase most of it (daily) in the market. Fez in the Maghrib, on the other hand, lies inland (and is) surrounded by cold hills. Its inhabitants can be observed to look sad and gloomy and to be too much concerned for the future. Although a man in Fez might have provisions of wheat stored, sufficient to last him for years, he always goes to the market early to buy his food for the day, because he is afraid to consume any of his hoarded food.

    If one pays attention to this sort of thing in the various zones and countries, the influence of the varying quality of the air upon the character (of the inhabitants) will become apparent. God is the Creator, the Knowing One. Al-Masudi undertook to investigate the reason for the levity, excitability, and emotionalism in Negroes, and attempted to explain it. However, he did no better than to report, on the authority of Galen and Ya'qub b. Ishaq al Kind!, that the reason is a weakness of their brains which results in a weakness of their intellect. This is an inconclusive and unproven statement. God guides whomever He wants to guide.
\end{quote}
Like Hippocrates and Aristotle, Ibn Khaldun argued that climate determines the character of a people, and that his climate was the best. He adds teleology: Arabs were superior because Allah had bestowed them with the optimal climate. The dominant position of Arabs was not by happenstance, but God's will.

Montesquieu\href{https://oll.libertyfund.org/titles/montesquieu-complete-works-vol-1-the-spirit-of-laws}{(1748, The Spirit of Laws, 1 XIV)} argues that the climate of France is best. He writes that
\begin{quote}
    [...] the temper of the mind and the passions of the heart are extremely different in different climates [...]
    
    People are therefore more vigorous in cold climates. [...] In cold countries they have very little sensibility for pleasure; in temperate countries they have more; in warm countries their sensibility is exquisite. [...]
    
    In northern climates, scarcely has the animal part of love a power of making itself felt. In temperate climates, love, attended by a thousand appendages, endeavours to please by things that have, at first, the appearance, though not the reality, of this passion. In warmer climates, it is liked for its own sake, it is the only cause of happiness, it is life itself.
\end{quote}
In other words, only the French have the right mix of strength and purpose, because only France has the right climate.

More than 200 years later, Ellsworth Huntington \href{https://www.questia.com/read/9043355/civilization-and-climate}{(1915, Civilization and Climate)} wrote
\begin{quote}
    Today a certain peculiar type of climate prevails wherever civilization is high. In the past, the same type seems to have prevailed wherever a great civilization arose.
\end{quote}
and
\begin{quote}
    In tropical countries weakness of will is unfortunately a quality displayed not only by the natives, but by a large proportion of the northerner sojourners. If manifests itself in many ways. Four of these, namely, lack of industry, an irascible temper, drunkenness, and sexual indulgence are particularly prominent, and may be taken as typical.
\end{quote}
At the same time, he argued that
\begin{quote}
    the effect of a diverse inheritance would last indefinitely
\end{quote}
in a thought experiment about "Teutons" and "negroes" moving to an empty country much like Egypt.\footnote{\citet{Bauval2011} argue for a strong influence of black Africans on early Egypt.} This racist and self-serving reasoning makes environmental determinism disreputable in my eyes and many of my contemporaries \citep[e.g., ][]{Peet1985, Donner2020}.

Lucien \citet{Febvre1922} pushed back against environmental determinism, emphasizing the social factors that typically dominate natural influences. Peter \citet{Frankopan2024} defends the same thesis, but with an additional century of data collection. Frankopan recounts how similar weather events and climate shifts led to different societal outcomes depending on economic and political circumstances\textemdash similar weather events can have big or small consequences, good or bad ones. He also shows how the same consequence, say the fall of an empire, was associated with very different weather events. Febvre and Frankopan argue against monocausalism.

These books range widely. Journal papers are more focused in their critique of environmental determinism. See, for example, \citet{Erickson1999} on the collapse of Andean societies, \citet{Coombes2005} on the demise of the Akkadian empire, \citet{Sessa2019} on the fall of the Western Roman one, or \citet{Slavin2016} on famine. These authors emphasize the many contributing factors and their complex interplay.

\begin{figure}
    \centering
    \caption{The temperature of the largest cities over time.}
    \label{fig:citytemp}
    \includegraphics[width=1.0\linewidth]{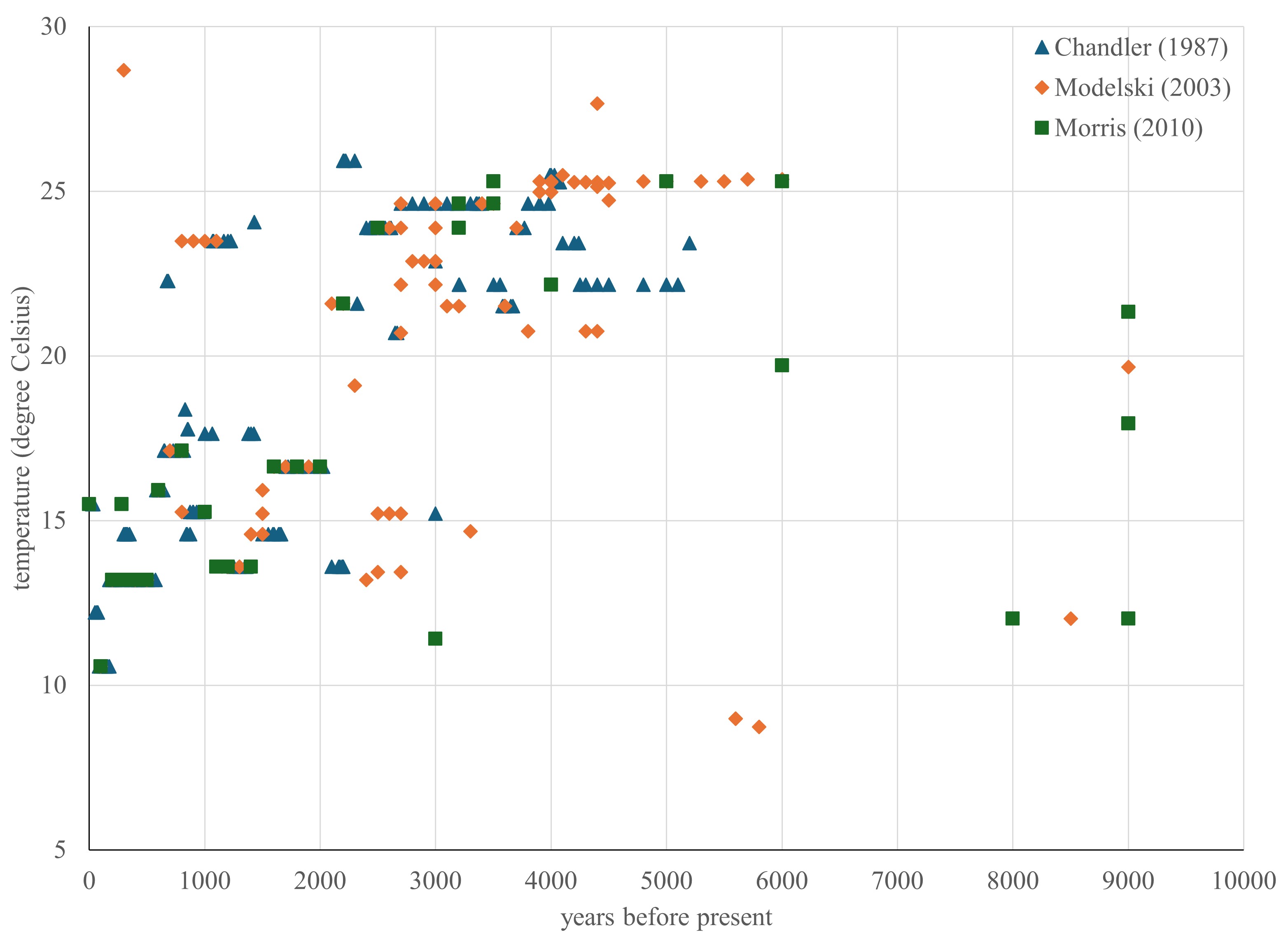}
    \caption*{\footnotesize The graph shows the average annual temperature between 1990 and 2020 at the location of the then-largest cities on Earth according to three different studies of city populations. Sources: Cities from \citet{Chandler1987}, \citet{Modelski2003} and \citet{Morris2010}. Climate data from \citet{Harris2020}.}
\end{figure}

The centres of military, political, and economic power have shifted considerably over time; their climates cannot all have been optimal. Figure \ref{fig:citytemp} illustrates this. The largest, presumably hegemonic cities varied in temperature between 8 and 28\celsius. There may be a reversal of fortune from an earlier dominance of the subtropics to a later dominance of the temperature zone \citep{acemoglu2002reversal}, but it is clear that the gods did not bestow the hegemon with the optimal climate\textemdash or maybe different gods have different views on the optimal climate.

\section{Climate determinism present}
\label{sc:present}
The above brief survey of climate determinism past would be an amusing example of the silly things people used to believe\textemdash would be, if climate determinism had disappeared from the discourse.

It has not. Jared \citet{Diamond1997} is probably the most prominent of current environmental determinists, but he is not alone \citep{Landes1998, Herbst2000, Vliert2008, Hausmann2009, Behringer2010, Morris2010}. Diamond argues that poor countries are poor because of unfavourable geography. The presence of domesticable species drives the emergence of agriculture. The orientation of continents determines how easily innovations, such as domesticated plants and animals, can spread\textemdash East-West implies large, contiguous areas of similar climates, North-South leads to fragmentation. Diamond uses these two factors to explain the emergence of the first centres of civilization and prosperity. Later, geographical barriers, such as rivers and mountain ranges, created conditions for political competition and innovation. This would explain the different histories of China and Europe.\footnote{But not India \citep{Weldon2025}. Weldon also argues that the \textit{Pax Mongolica} enabled Europe to catch up with Chinese technology, while later political instability along the Silk Road led to the exploration of sea routes to Asia.} There is limited empirical support for Diamond's thesis \citep{Olsson2005}, but the book was roundly criticized \citep[e.g.,][]{Blaut1999, Sluyter2003}.

\citet{Diamond2005} doubles down. While the 1997 book argues that favourable geography explains the rise of civilizations, the 2005 book argues that unfavourable geography explains their collapse. \citet{Kemp2025} and particularly \citet{Cooper2025} take a more nuanced approach, discussing the multi-causal nature of the fall of past civilizations.

Climate determinism can also be found in parts of the environmental movement. The organizations \href{https://extinctionrebellion.uk/}{\textit{Extinction Rebellion}} and the \href{https://letztegeneration.org/}{\textit{Last Generation}}, for instance, argue that environmental change will lead to the extinction of humankind. The former organization aims to ``ensure the future of all life on Earth'' and ``\emph{our} survival'' (emphasis added). The latter argues that ``we are the last generation who may be able to stop the complete collapse of Earth''. See \citet{Gunningham2019, Martiskainen2020, Westwell2020} and \citet{Moor2021} for a discussion of these organizations. \citet{Tol2017struct} argues that casting climate change in the familiar apocalyptic narrative is a good strategy for maximizing donations. Regardless of whether it is a cynical ploy or a firmly held belief, if climate change is powerful enough to threaten an advanced civilization and a ubiquitous generalist, then surely climate must have determined their current condition.

Environmental scientists too have flirted with catastrophism. ``Kate Marvel has seen the world end'' is the opening line of the blurb of her otherwise marvelous book \citep{Marvel2025}. She is not alone; other scientists have raised the prospect of the end of the world \citep{Karplus1992, Levy2005, Hansen2007, Wallace2019, Galbraith2021, Kemp2022, Davidson2024}.\footnote{Others note that the approaching end times are a constant in human culture, although the nature and cause of the coming apocalypse differ over space and time \citep{Pearson2006, Landes2011}.}

Economists did not accept the premise of environmental determinism. Adam \citet{Smith1776} acknowledged that rivers and ports are important for navigation, but argued that competition and specialization determine the wealth of nations. The first model of economic growth \citep{Harrod1939, Domar1946, Domar1947} was limited to capital accumulation. \citet{Lewis1954} focused on labour, \citet{Boserup1965} on intensification. \citet{Solow1956} found that technological progress was the key driver, a view reinforced by \citet{Cohen1990, Romer1990, Aghion1992, Mankiw1992, Mokyr1992} and \citet{Audretsch1996}. \citet{North1990} argues that institutions are key, as do \citet{Mauro1995} and \citet{Besley2009}. In short, economists who study economic growth and development have consistently argued that other factors dominate the impact of climate and geography.

Jeffrey Sachs revived economists' interest in the impact of climate and geography on economic development \citep[][see also \citet{Masters2001, Bloom2003, Olsson2005, Nordhaus2006, Dell2009, Meierrieks2024, Lakhtakiafc, Tolfc, Kumar2025}]{Bloom1998, Gallup1999, Sachs2001, Sachs2002, Sachs2003}. His argument centres on the burden of vector-borne diseases, particularly malaria, on labour productivity, both directly and indirectly through cognitive development and fertility choices.

Sachs emphasizes that these diseases can and should be prevented and is therefore not an environmental determinist in the standard sense of the word. Nonetheless, in response, people argued that only institutions matter for economic development \citep{acemoglu2002reversal, Easterly2003, Rodrik2004}. \citet[][but see \citet{Albouy2012}]{Acemoglu2001} provided a twist to this institutional determinism: The disease environment shaped institutions. This argument was later reinforced by \citet{Alsan2015} and extended to the deeper past by \citet{Galor2022} and \citet{Olsson2024}. The core argument is peculiar: Something from hundreds or thousands of years ago created an immutable feature in a nation's culture or institutions. \citet{Conley2025} show that many of these empirical results vanish if spatial autocorrelation is properly controlled for.

\begin{figure}
    \centering
    \caption{Range of per capita income by temperature.}
    \label{fig:incometemp}
    \includegraphics[width=1.0\linewidth]{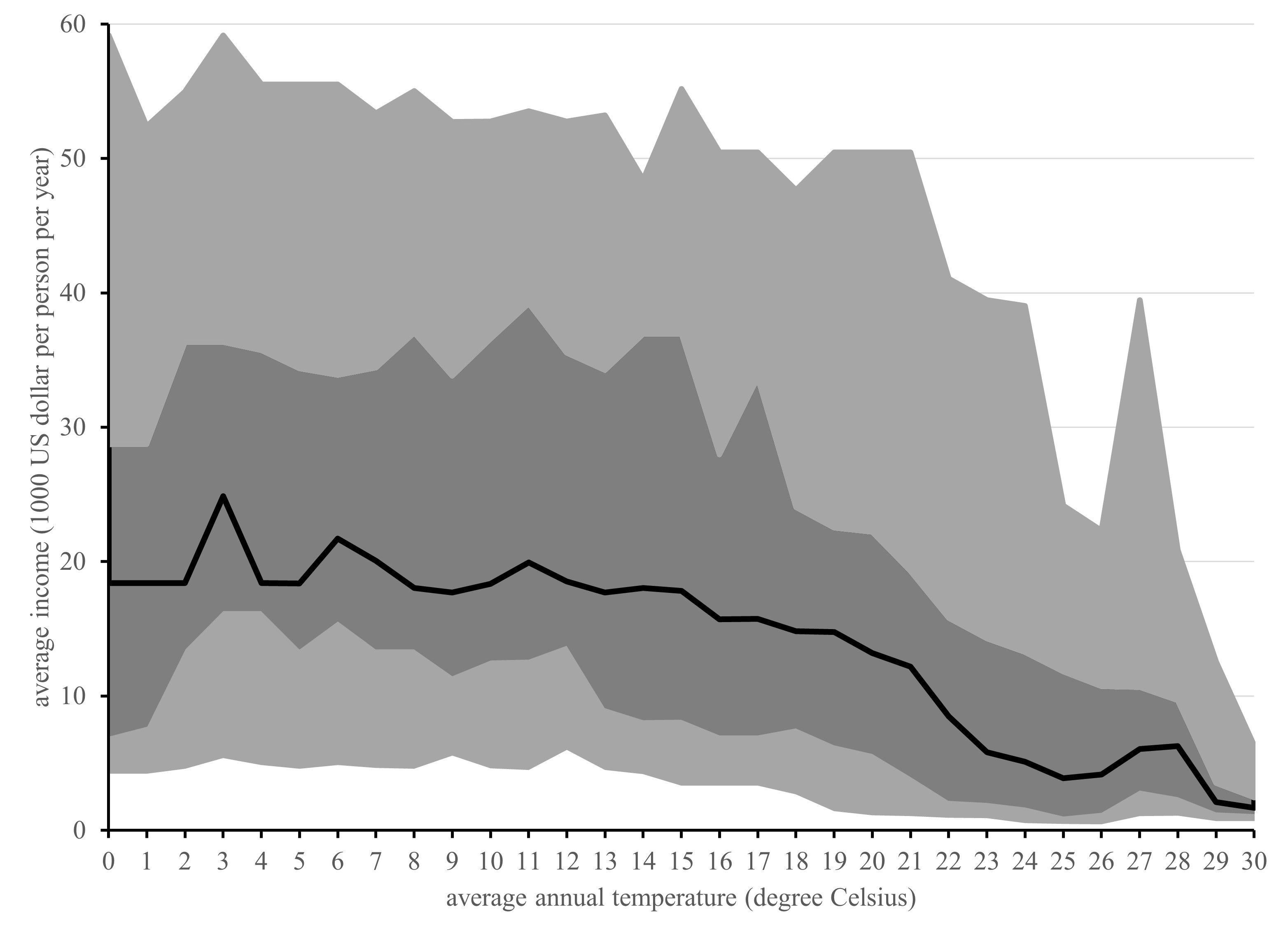}
    \caption*{\footnotesize The graph shows median per capita income (black line), the range between the 25th and 75th percentile (dark gray area), and the range between the 5th and 95th percentile (light gray area). Source: Data from \citet{Tol2024niche}.}
\end{figure}

The correlation between income and temperature is weak. Figure \ref{fig:incometemp} plots the 5th, 25th, 50th, 75th and 95th percentiles of per capita income against the annual mean temperature. I use \emph{average} income for the first subnational level (states, provinces, counties, oblasts ...), uniformly project these on a 1/6\texttimes 1.6\degree{} grid, and then reaggregate to 1\celsius{} temperature bins. The median income tends to be lower in hotter places, but the spread is large. There are poor and rich regions for every temperature, except at the highest temperatures, where poverty is more uniform.

A better understanding of the causes of development helps to improve economic policy, but it is not immediately relevant to the discussion on future climate change. Endemic malaria may have determined the behaviour of European colonists, but that was then. Some studies instead look at the recent past.

\citet[][cf. \citet{Barker2023}]{Dell2012} find a persistent, negative effect of high temperature on economic growth in poor countries. Specifically, a 1\celsius{} temperature rise would reduce annual income growth by 1.4 percentage points. This is a large effect; a factor 4 over a century. The world has warmed by more than 1\celsius{} since pre-industrial times. If Dell is right, poor countries are poor because they are hot; poor countries will get poorer if it gets hotter still.

\citet[][cf. \citet{Barker2024BHM}]{Burke2015}\footnote{They repeat the analysis in \citet{Burke2018nat} and \citet[][see also \citet{Rosen2019}]{Diffenbaugh2019}. Ten years and much ignored criticism later, \citet{Burke2025} re-estimate the same model.} use a different specification: Where Dell and coauthors assume that \emph{poor} countries are more vulnerable to weather shocks, Burke and colleagues assume that \emph{hot} countries are. This specification was adopted by \citet{Pretis2018, Henseler2019, Acevedo2020, Damania2020, Kikstra2021, Callahan2025} and \citet{ Neal2025}\footnote{\citet{Ahmadi2025} and \citet{Apergis2025} assume uniform vulnerability.} The distinction is subtle in-sample, as poor countries tend to be hot, but the difference is substantial in a richer (Dell: positive) and hotter (Burke: negative) future.

\begin{figure}
    \centering
    \caption{Observed and counterfactual growth rates.}
    \label{fig:growth}
    \includegraphics[width=1.0\linewidth]{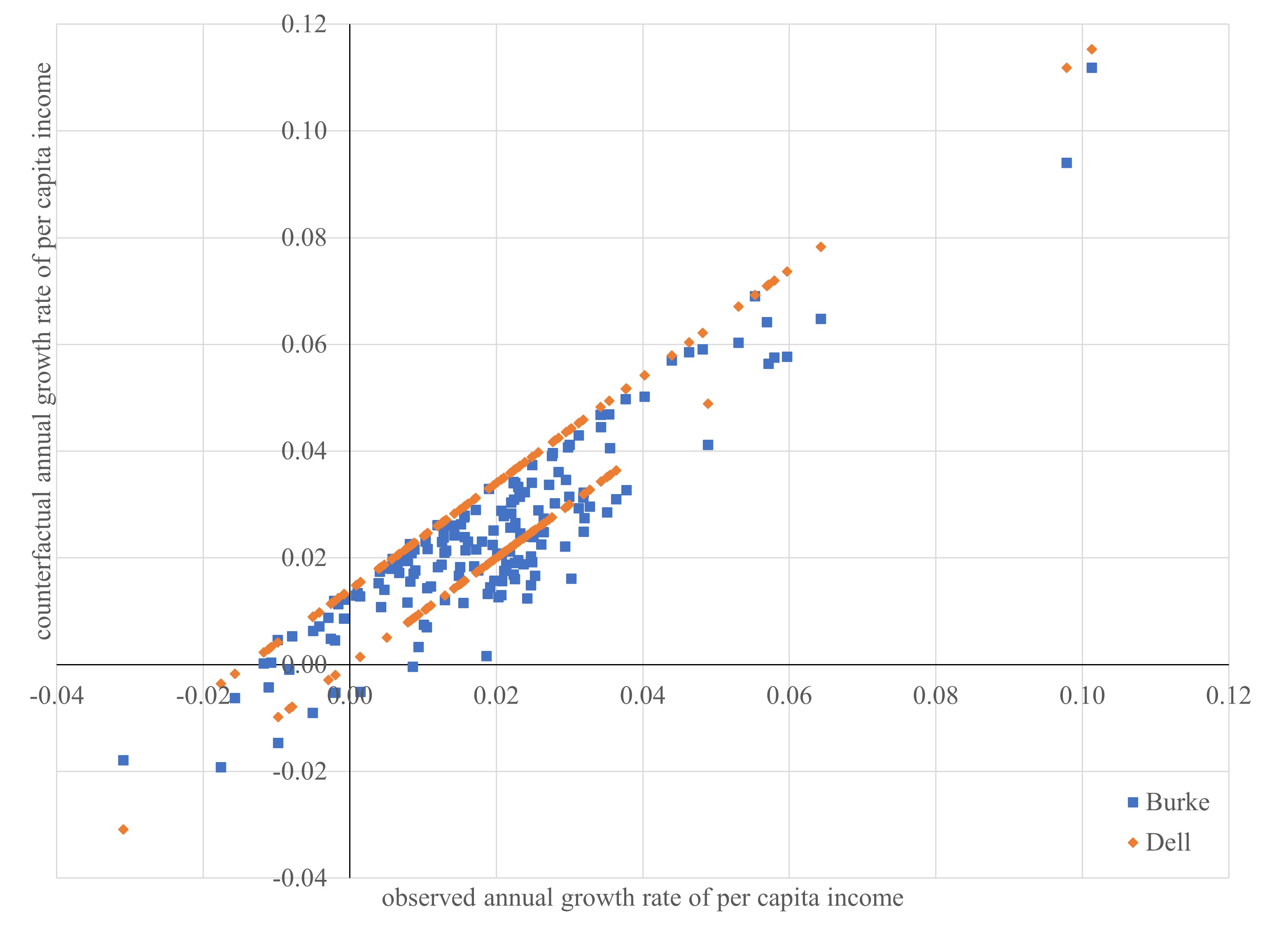}
    \caption*{\footnotesize The graph plots the observed growth rate of per capita income against the counterfactual growth rates had the world not warmed by 1\celsius. Source: Data from \citet{Burke2015}; parameters from \citet{Dell2012} and \citet{Burke2015}.}
\end{figure}

Both Burke and Dell find large effects. Figure \ref{fig:growth} plots the observed average annual growth rate against the counterfactual had the world not warmed by 1\celsius. For poor countries, Burke's estimates are generally smaller than Dell's. For rich countries, Burke attributes a large part of observed growth to climate change. Russia's economy, for instance, would have been stagnant had it not been for global warming, while per capita income in the UK would have grown at 1.6\% rather 2.0\%. On the other side of the temperature spectrum, Nigeria's per capita income would have grown at 2.6\% per year rather than 1.3\%, and Ghana's at 2.2\% rather than 0.9\%. The implication is, of course, that at the start of Industrial Revolution when it was colder, economic growth was slow in the UK and fast in Nigeria. This flies in the face of historical facts.

Projected impacts are large too. For instance, \citet{Burke2015} reports that end-of-century average per capita income in the richest country would be four times as large with climate change than without. In the poorest country, climate change would take away 90\% of income.\footnote{The catastrophic impact of \citet{Weitzman2009} is different: The structural uncertainty about climate change is so large that it violates the axioms of cost-benefit analysis \citep{Tol2003, Anthoff2014}.}

Note that Figure \ref{fig:growth}, like Dell and Burke, compares each country to itself. This obscures the true effect sizes. According to Burke's central estimate, any economy with an average temperature above 26\celsius{} would have shrunk. Country-fixed effects compensate for this\textemdash within sample. According to Dell, Indonesia faces a climate penalty of 35 percentage points in its annual growth rate, and India 36. Again, country-fixed effects make up the difference between the climate effect and the observed growth rate. Large effect sizes offset by fixed effects spell trouble. Indeed, \citet{Newell2021} find that these models do not extrapolate well. A spatial cross-validation that omits the observations for certain countries cannot predict said observations because the country-fixed effects are set to zero. At the same time, time-fixed effects compensate for the trend in the global temperature \citep{Berg2024}. Therefore, temporal cross-validation, omitting certain periods, also fails \citep{Newell2021}.

\citet{Bilal2024} make a similar mistake. Dell and Burke use country fixed effects. This means trouble for the economy if the temperature deviates from its long-term average. Bilal and K\"{a}nzig instead use local projections \citep{Jorda2005}. This means trouble for the economy if the temperature deviates from its trend. This is a more plausible representation of human behaviour\textemdash fixed effects are often called dummies, so local projections should be called smarties. However, like fixed effects, local projections are only available in-sample. In their forecasts, the projected temperature is set to zero and the impact of 1\celsius{} warming is a drop in economic activity of 30\%. This effect size would make the ancient environmental determinists blush. 

There is a structural problem with regressions like those of Dell and Burke. The dependent variable is a rate, the explanatory variable a level. They regress a flow on a stock. The left-hand side is stationary, the right-hand side non-stationary. This explains why these estimates are not robust \citep{Newell2021, Tol2024EnPol}. For instance, \citet{Harding2025} and \citet{Tolfc} recast the models of Dell and Burke in a macroeconomic growth framework and find much smaller effects. \citet{Lakhtakiafc} show that adding lags turns a permanent effect into a transient one. As another example, \citet{Berg2024} split the temperature record into a global and a national component to find that \emph{rich} countries are more vulnerable to weather shocks. 

\citet{Lemoine2016, Letta2018, Kahn2019} and \citet{Meierrieks2024} therefore regress the economic growth rate on the \emph{change} in temperature. The estimated impact of warming is two orders of magnitude smaller \citep{Newell2021, Tol2024EnPol}.

\citet{Kalkuhl2020, Kotz2021, Kotz2022}, \citet[][cf. \citet{Barker2024}]{Kiley2024} and \citet[][cf. \citet{Bearpark2025} and \citet{Schotz2025}]{Kotz2024} take a different approach. They regress growth on the change in temperature and the interaction between the change in temperature and its level. They thus make the same mistakes as Dell and Burke. They omit the level of temperature from the regression, because it would be constant. (It is not.) A constant temperature level implies that the interaction effect cannot be interpreted and induces a correlation between the variable of interest and the time and location fixed effects.

The legitimate concerns about the validity and robustness of Dell- and Burke-like models appear not to have dented their popularity. \citet{Dell2012} has been cited 3,058 times, \citet{Burke2015} 3,355 times (Google Scholar, 29 August 2025). Both models imply climate determinism: The economic growth rate depends on the temperature. There is a structural impediment to economic growth in hot countries, or perhaps in hot and poor countries.

Models like these appear to perform well in-sample because the climate effect is offset by unexplained factors that are anti-correlated with temperature. These fixed effects\textemdash a euphemism for something we do not understand, a dummy really\textemdash obscure the nature of these models. Climate determinism is veiled but real.

\section{Climate determinism yet-to-come}
\label{sc:future}
Climate affected past development and hence the current distribution of income, but this impact was not profound. Similarly, climate change will affect future economic growth, but is unlikely to change its course. Can climate policy?

At first sight, the answer is no. Climate policy will have a minor impact on economic development. Successive assessments of the costs of greenhouse gas emission reduction point to modest costs \citep{Weyant1993, Clarke2009, Riahi2022IPCC}. There are two reasons for this. First, energy is a small part of the economy. Therefore, a gradual transition from cheaper fuels to more expensive fuels cannot have a major impact. Moreover, renewables are now cheaper than fossil fuels in an ever-growing part of the energy market.

The second reason is that most studies assume that climate policy will be first-best\textemdash that emissions will be reduced at the lowest possible cost. This is a tall assumption. Actual climate policy is far from perfect. A recent comparison of \textit{ex ante} to \textit{ex post} cost estimates found that the econometric studies are close to the more pessimistic end of the range of cost estimates from calibrated models \citep{Tol2023EAP}. That said, these estimates still point to a modest impact of \emph{current} climate policy on economic development.

\begin{figure}
    \centering
    \caption{The Leviathan tax and current and projected carbon prices.}
    \label{fig:leviathan}
    \includegraphics[width=1.0\linewidth]{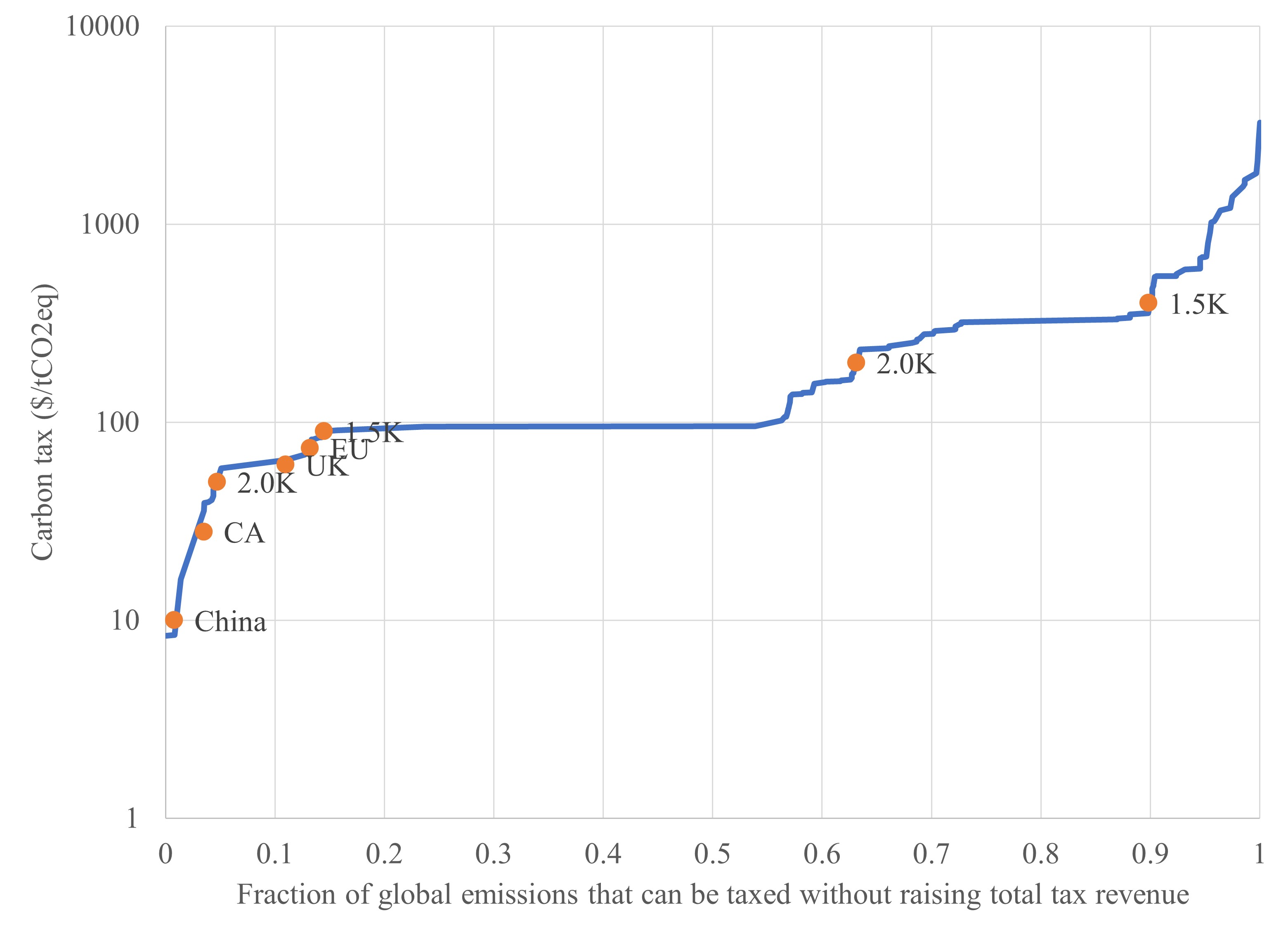}
    \caption*{\footnotesize The Leviathan tax is for 2019. Current carbon prices are the most recent prices for CO\textsubscript{2} emission permits. Projected carbon prices are for 2030; I show both the median and the maximum reported. Sources: \href{https://data.worldbank.org/}{World Bank Open Data}; \href{https://icapcarbonaction.com/en/ets-prices}{ICAP}; \citet{Riahi2022IPCC}.}
\end{figure}

What about \emph{future} climate policy? Figure \ref{fig:leviathan} compares the Leviathan tax to current and projected carbon prices. The Leviathan tax is the carbon tax that would allow for the abolition of all other taxes without shrinking (or growing) the public sector \citep{Tol2012CCL}. In 2019, it ranged from \$8/tCO\textsubscript{2} in the Central African Republic and Iraq to \$3,263/tCO\textsubscript{2} in Sweden. Current prices for carbon dioxide emission permits range widely, from \$6/tCO\textsubscript{2} in South Korea (not shown) and \$10/tCO\textsubscript{2} in China to \$74/tCO\textsubscript{2} in the European Union. The EU carbon price would imply a substantial fiscal challenge in India and China, where the Leviathan tax is \$95/tCO\textsubscript{2} and \$96/tCO\textsubscript{2}, respectively. If permits were auctioned, these countries could reduce their tax rates by some three-quarters. The IPCC \citep{Riahi2022IPCC} reports that the median model requires a global carbon tax of about \$50 (90)/tCO\textsubscript{2} by 2030 to meet the 2.0 (1.5)\celsius{} target of the 2015 Paris Agreement, while the most pessimistic model needs \$200 (400)/tCO\textsubscript{2}. The highest number equals the 90th percentile of carbon dioxide emissions.

Climate policy would reduce the carbon intensity of the economy and so increase the Leviathan tax\textemdash which also tends to fall over time with improvements in energy efficiency and the increasing competitiveness of renewable energy. Yet, stringent climate policy would make it a major factor in fiscal policy in many middle-income countries, and a dominant factor in low-income ones.

\citet{Tol2023EAP} reviews the IPCC emission reduction scenarios \citep{Riahi2022IPCC}. For the model with the highest skill in backcasting climate policy, for climate policy in line with the Paris Agreement, global carbon tax revenue would be 19\% by 2040\textemdash global tax revenue was 24\% in 2019. The subsidy for removing carbon dioxide from the atmosphere would amount to 15\% of GDP. As always, the global average hides diversity. Some 98\% of the economy in Russia would be emission removal.

Should this future come to pass, it would be rightly called climate determinism. However, unlike climate and climate change, climate policy is not imposed\textemdash it is (largely) self-imposed. It is hard to imagine any government implementing a climate policy that would have a drastic impact on fiscal policy. Indeed, Figure \ref{fig:leviathan} shows that current climate policy is far removed from its domestic upper limit.

The same is true for the export of emission reduction credits. Assuming a credit price of \$10/tCO\textsubscript{2}\footnote{Source: \href{https://alliedoffsets.com/pricing-activity/}{Allied Offsets}.}, the total carbon offset market is worth 0.004\% of global total economic output. More pertinently, the export of carbon credits is smaller than 0.1\% of GDP in all but nine countries: Belize (0.2\%), Cambodia (0.3\%), Central African Republic (0.2\%), Democratic Republic of the Congo (0.2\%), Guyana (0.9\%), Malawi (0.4\%), Rwanda (0.2\%), Zambia (0.2\%), and Zimbabwe (0.2\%). These are modest numbers.

It may, of course, happen that the demand for carbon credits from, say, Europe grows so large that it would distort, maybe even dominate, the economy of a small country. This is less likely to happen with a large country or a large number of small countries, as there are political limits on the value of imports by government fiat. That said, the prospect of a carbon offset resource curse, with negative ramifications for the structure of the economy and, perhaps, political institutions \citep{Ploeg2011JEL, Ross2015}, is real for some small countries.

\section{Discussion and Conclusion}
\label{sc:conclude}
Environmental determinism in the past can be explained by scholars seeking a divine explanation for the political and economic success of their benefactors. Its demise is readily understood: Evidence accumulated on the diversity of living conditions over space and over time, monocausal theories of power were replaced by more nuanced accounts, and the supernatural made way for human agency.

Current environmental determinism is harder to explain, at least in the scholarly literature, since it flies in the face of so much evidence. The popular literature exploits the public desire for simple narratives and scary stories. \citet{Tol2025anyas} documents something of a bidding war that is driving up estimates of the social cost of carbon. There are too few studies of the total economic impact of climate change to test this hypothesis but a similar mechanism may be at play. A widespread unease with earlier estimates and their implications for climate policy plus the career opportunities that beckon for showing a Nobel laureate wrong may have led to a suspension of the normal standard of econometric rigour\textemdash and many of these economics papers were published in extradisciplinary journals.

In the future, climate policy would dominate the economy if politicians make good on their rhetoric. My interpretation is that they therefore will not. Across Europe, we indeed see policymakers retreating from their predecessors' overly ambitious promises on greenhouse gas abatement. The temptation to import emission reduction credits will not disappear. Such imports will be small relative to the importing economy but may be large for the exporter, potentially dominating the economy and overwhelming institutions.

\bibliography{master}

\end{document}